\documentclass[fleqn,usenatbib]{mnras}

\usepackage{newtxtext,newtxmath}
\usepackage[T1]{fontenc}

\DeclareRobustCommand{\VAN}[3]{#2}
\let\VANthebibliography\thebibliography
\def\thebibliography{\DeclareRobustCommand{\VAN}[3]{##3}\VANthebibliography}

\usepackage{graphicx}	% Including figure files
\usepackage{amsmath}	% Advanced maths commands
\title[And XIX - a detailed star formation history]{A detailed star formation history for the extremely diffuse Andromeda XIX dwarf galaxy}

\author[Authors]{
Michelle L. M. Collins$^{1}$\thanks{E-mail: m.collins@surrey.ac.uk.uk (MLMC)},
Benjamin F. Williams$^{2}$,
Erik J. Tollerud$^{3}$,
Eduardo Balbinot$^{4}$,
\newauthor Karoline M. Gilbert$^{3,5}$  and Andrew Dolphin$^{6,7}$
\\
$^{1}$ Physics Department, University of Surrey, Guildford, GU2 7XH, UK\\
$^{2}$ Department of Astronomy, Box 351580, University of Washington, Seattle, WA 98195, USA\\
$^{3}$ Space Telescope Science Institute,
$^{4}$ Kapteyn Astronomical Institute, University of Groningen, Postbus 800, NL-9700AV Groningen, the Netherlands\\
$^{5}$ The William H. Miller III Department of Physics \& Astronomy, Bloomberg Center for Physics and Astronomy, John Hopkins University, \\ 3400 N. Charles Street, Baltimore, MD 21218\\
$^{6}$ Steward Observatory, University of Arizona, Tucson, AZ 85719, USA\\
$^{7}$ Raytheon, Tucson, AZ 85734, USA
}

\date{Accepted XXX. Received YYY; in original form ZZZ}

\pubyear{2020}

\begin{document}
\label{firstpage}
\pagerange{\pageref{firstpage}--\pageref{lastpage}}
\maketitle

\begin{abstract}
We present deep imaging of the ultra-diffuse Andromeda XIX dwarf galaxy from the Advance Camera for Surveys on the \emph{Hubble Space Telescope} which resolves its stellar populations to below the oldest main sequence turn-off. We derive a full star formation history for the galaxy using MATCH, and find no evidence of star formation in the past 8~Gyr. We calculate a quenching time of $\tau_{90}=9.7\pm0.2$~Gyr, suggesting Andromeda~XIX ceased forming stars very early on. This early quenching, combined with its extremely large half-light radius, low density dark matter halo and lower than expected metallicity make it a unique galaxy within the Local Group and raises questions about how it formed. The early quenching time allows us to rule out feedback from bursty star formation as a means to explain its diffuse stellar population and low density dark matter halo. We find that the extended stellar population, low density halo and star formation could be explained by either tidal interactions (such as tidal shocking) or by late dry mergers, with the latter also explaining its low metallicity. Proper motions and detailed abundances would allow us to distinguish between these two scenarios.
\end{abstract}


%%%%%%%%%%%%%%%%%%%%%%%%%%%%%%%%%%%%%%%%%%%%%%%%%%

%%%%%%%%%%%%%%%%% BODY OF PAPER %%%%%%%%%%%%%%%%%%

\section{Introduction}

Since the moment Andromeda (And) XIX was first discovered in the Pan-Andromeda Archaeological Survey (PAndAS, \citealt{mcconnachie08}), the M31 satellite galaxy has intrigued astronomers. The dwarf spheroidal (dSph) has a luminosity of $L=9.4\times10^5\,{\rm L}_\odot$, placing it near the boundary of `classical' and `ultra-faint' dSph companions of the Milky Way. However, with a half light radius of $r_{\rm half}= 3357^{+816}_{-465}\,{\rm pc}$ \citep{martin16c,savino22}, it is approximately a factor of 10 times more extended than similarly luminous dwarf galaxies in the Local Group. Its surface brightness is orders of magnitude lower than so-called ``ultra diffuse'' galaxies (UDGs, \citealt{vandokkum15a}), comparable only to the similarly extreme Antlia 2 dwarf galaxy in the Milky Way \citep{torrealba18}. In addition to their unusual structural properties, both And~XIX and Antlia 2 have lower central dark matter densities than predicted by the standard $\Lambda$ Cold Dark Matter ($\Lambda$CDM) cosmological model  \citep{torrealba18,collins20}.

This combination of properties presents a puzzle for understanding the formation of And~XIX. Why is it so much more extended than similarly bright galaxies? A few mechanisms have been proposed to understand its morphology (and that of Antlia~2). They could have experienced significant tidal interactions with their large spiral hosts, which reshaped their structure and lowered their density \citep[e.g.][]{ogiya18,amorisco19a,collins20,montes20,ji21,jackson21,maccio21,ogiya22,moreno22}. In deep imaging from PAndAS, And XIX appears extended along its major axis, with low surface brightness features that could be indicative of tidal interaction \citep{mcconnachie08, collins20}. Another explanation could be that they experienced powerful, bursty star formation, which has redistributed their dark and stellar material \citep[e.g.][]{read19a,torrealba18}. Or, could they be the product of an early galactic merger \citep[e.g.][]{silk19,rey19,shin20}?

\citet{collins20} presented spectroscopic data for over 100 stellar members of And~XIX to try to address this issue. They found that And~XIX has a lower density dark matter halo than expected, and with a mean metallicity of [Fe/H]$=-2.07\pm0.02$, it is a metal-poor outlier from the \citet{kirby13b} luminosity-metallicity relation. They also found evidence for a slight velocity gradient across its major axis, and a dynamically cold 'spot' on its south-eastern extent. These findings led them to conclude that And~XIX may not be in dynamical equilibrium, but did not allow them to determine conclusively why it looks as extreme as it does.

Here, we present further insight into the evolution of And~XIX. Using the Hubble Space Telescope (HST) we have acquired exquisite deep imaging of And~XIX to below its oldest main sequence turn-off. From this, we can derive its star formation history (SFH) allowing us to test whether bursty star formation has been important in its evolution, and search for clues as to its true nature. We present the data and our methods for measuring the SFH in \S~\ref{sec:method}, our results follow in \S~\ref{sec:results}, and we discuss their implication in \S~\ref{sec:disc}, before concluding in \S~\ref{sec:conclusion}.

\section{Method}
\label{sec:method}

\subsection{The data}

\begin{figure*}
	\includegraphics[width=\columnwidth]{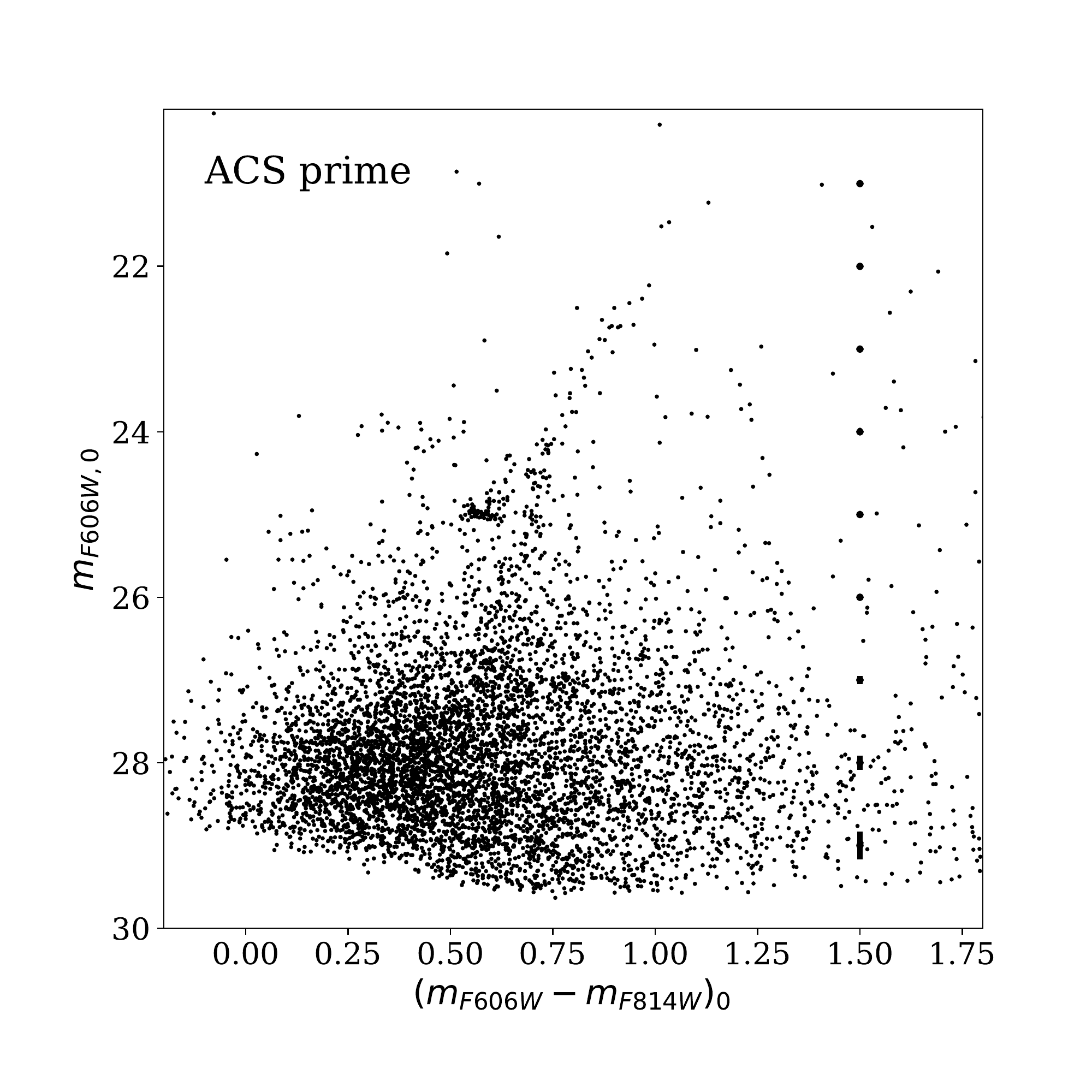}
	\includegraphics[width=\columnwidth]{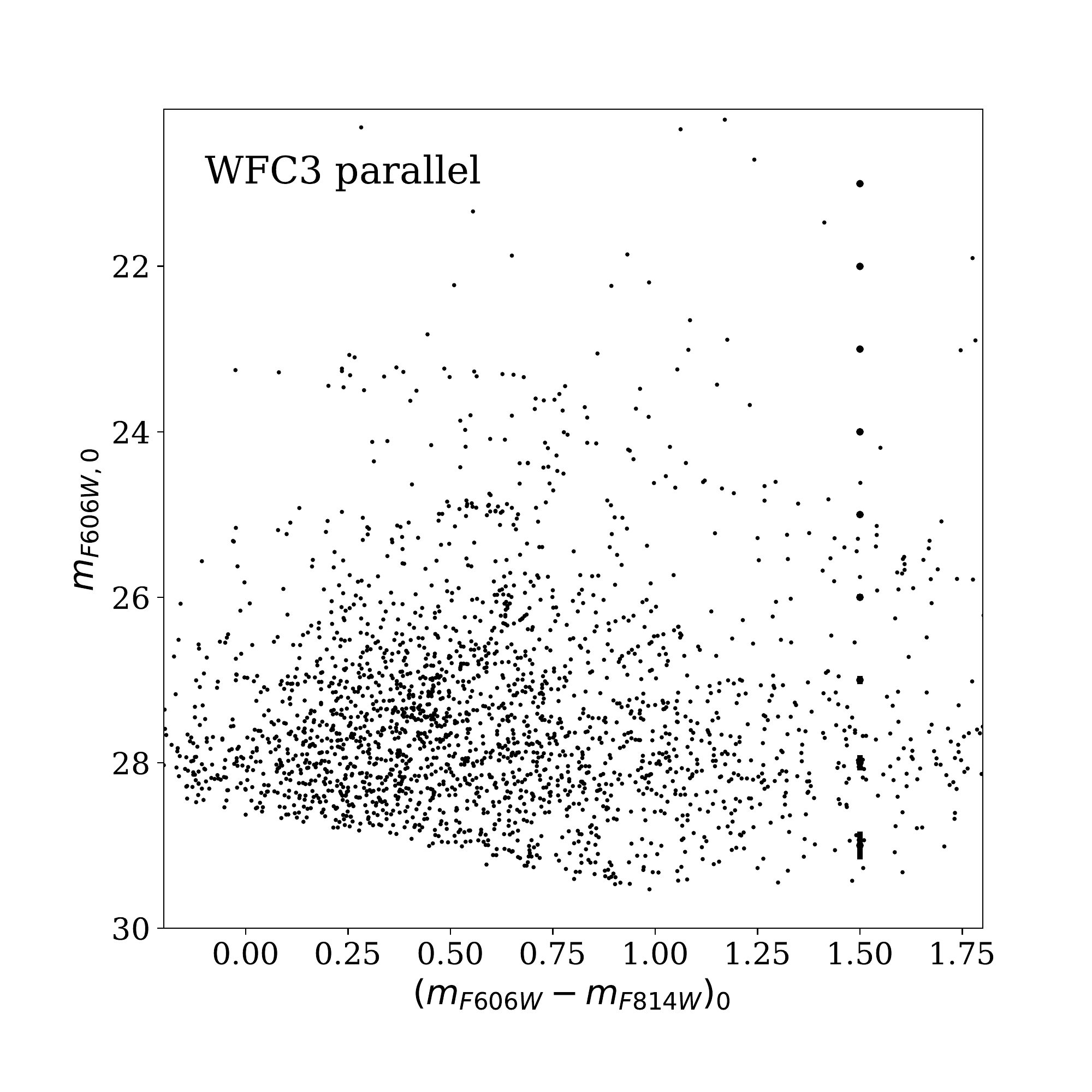}
\caption{{\bf Left:} The extinction corrected CMD for our ACS prime observations for And~XIX. All sources passing our sharpness and crowding conditions are plotted. The horizontal branch is clearly seen at $m_{\rm F606W}\sim25$. Representative error bars are plotted on the right hand side. {\bf Right:} The same but for our WFC3 parallel field. some evidence for the stellar population of And XIX can be seen, but the few stars make a clear detection of the dwarf difficult in this field.  }
    \label{fig:hst}
\end{figure*}

% \begin{figure}
%     \includegraphics[width=\columnwidth]{A19_LF.pdf}
% \caption{A luminosity function for And~XIX in the pseudo V band. The peak of the HB is seen at $m_V=24.8\pm0.02$, and is highlighted with a dash line.  }
%     \label{fig:hb}
% \end{figure}

The observations for this paper were taken through HST-GO-15302 (PI Collins), and were acquired between 23-28 Sept 2018. We observed the centre of And~XIX for 16 orbits using the Advanced Camera for Surveys (ACS), splitting the run into 2 orbit visits. Each visit allowed exposure times of 2460~s with the F606W filter and 2598~s with the F814W filter. We took simultaneous parallel observations in the south-east of And~XIX using the Wide Field Camera 3 (WFC3), again split into exposures of 2638~s with the F606W filter and 2758~s with the F814W filter. We summarise these observations in table~\ref{tab:obs}. The total exposure times were designed to allow us to reach a signal-to-noise of $S/N>8$ for the oldest main sequence turn-off (oMSTO) in And~XIX (estimated as 28.1 mag in F606W and 27.7 mag in F814W).

\begin{table*}
	\centering
	\caption{Details of observations}
	\label{tab:obs}
	\begin{tabular}{lcccc} % four columns, alignment for each
		\hline
		Instrument & RA (hh:mm:ss) & dec (dd:mm:ss) & Exposure time $F606W$ & Exposure time $F814W$ \\
		\hline
		\hline
		ACS/WFC (Prime) & $00:19:35.4$ & $+35:04:20$ & 21,104~s & 22,064~s\\
		WFC3/UVIS (Parallel)  & $00:19:16.4$ & $+34:59:57$ & 21,104~s & 22,064~s\\
		\hline
		\end{tabular}
\end{table*}

\begin{table}
	\centering
	\caption{Properties of And~XIX}
	\label{tab:props}
	\begin{tabular}{lc} % four columns, alignment for each
		\hline
		Property & \\
		\hline
		\hline
		RA (J2000) & $00:19:34.5^a$ \\
		dec (J2000) &$+35:02:41^a$ \\
		$m_V$ & $14.5\pm0.3^a$\\
		$\mu_{V,0}$ (mag arcsec$^{-2}$ & $29.3\pm0.4^a$\\
		$D$ (kpc) & $812.8^{+24,c}_{-39}$\\
		$M_V$ &$-10.1\pm0.3^c$ \\
		$r_{\rm half}$ (arcmin) & $14.2^{+3.4,a}_{-1.9}$\\
		$r_{\rm half}$ (pc) & $3357^{+816,c}_{-465}$\\
        $\tau_{50} (Gyr)$ & $13.5\pm0.1^b$\\
		$\tau_{90}$ & $9.7\pm0.2^b$ \\ 
		${\rm [Fe/H]}$ & $-2.07\pm0.02^d$\\
		\hline
		$^a$ - \citet{martin16c}\\
		$^b$ - This work\\
		$^c$ -\citet{savino22}\\
		$^d$ -\citet{collins20}
		\end{tabular}
\end{table}

Given the extremely diffuse nature of And~XIX, we were only able to confidently detect the stellar populations of the galaxy in our central (highest surface brightness) ACS field. In fig.~\ref{fig:hst} we show the extinction corrected colour magnitude diagrams (CMDs) for both fields (computed using extinction and reddening values from \citealt{schlafly11}, using the \citealt{cardelli89} extinction curve and $R_V=3.1$). We include all point-like sources that do not suffer from crowding. The central ACS field shows a clear RGB and HB population. In the parallel WFC3 field, hints of both can be seen, but not enough stars are available for a quantitative analysis. As such, we were able to use the WFC3 data as a conservative background population for our MATCH analysis described below. 

Using the bias, flat-field, and image distortion corrected images from the STScI pipeline, we applied the PSF-fitting photometry package, {\sc Dolphot} \citep{dolphin00,dolphin16} to obtain photometry of the resolved stars. We applied the same pipeline steps for measuring resolved stellar photometry as used for the Panchromatic Hubble Andromeda Treasury \citep{williams14}, with a few updates.  The updates include the use of TinyTim point spread functions \citep{krist2011} and the use of ACS/WFC and WFC3/UVIS images that were corrected for charge transfer efficiency degradation \citep{andersen2010}.   We also performed 10$^5$ artificial star tests (ASTs), in which a star of known colour and brightness was injected into the original images and the photometry was remeasured.  The output of these ASTs provided a quantitative estimate of our bias, uncertainty, and completeness as a function of colour and magnitude. Our 90\% completeness depths are $m_{\rm F606W}=29.57$ and $m_{\rm F814W}=28.84$, meaning we probe approximately one magnitude deeper than the oMSTO in both bands.

\subsection{Measuring the star formation history of And~XIX}

Our extremely deep data are 90\% complete to approximately 1 magnitude below the oMSTO for And~XIX, allowing us to reliably probe the oldest stellar populations of the galaxy. Such high fidelity data allow us to break the age-metallicity degeneracy which plagues shallower data. At present, only 6 other M31 dwarf spheroidals have had SFHs from similarly deep data published \citep{weisz14a,skillman17}. A further 20 have had their SFH measured from shallower imaging reaching just below the horizontal branch (HB) in \citet{weisz19a}. Based on comparisons for the 6 objects for which both shallow and deep data are present, \citet{weisz19a} showed that the results are consistent, however the deeper data allow a more precise measure of the ancient SFH. As such, by measuring a detailed SFH for And~XIX, we can compare its evolution over cosmic time with the remaining M31 satellites, and place its evolution into a wider cosmological context.

\begin{figure*}
	\includegraphics[width=\columnwidth]{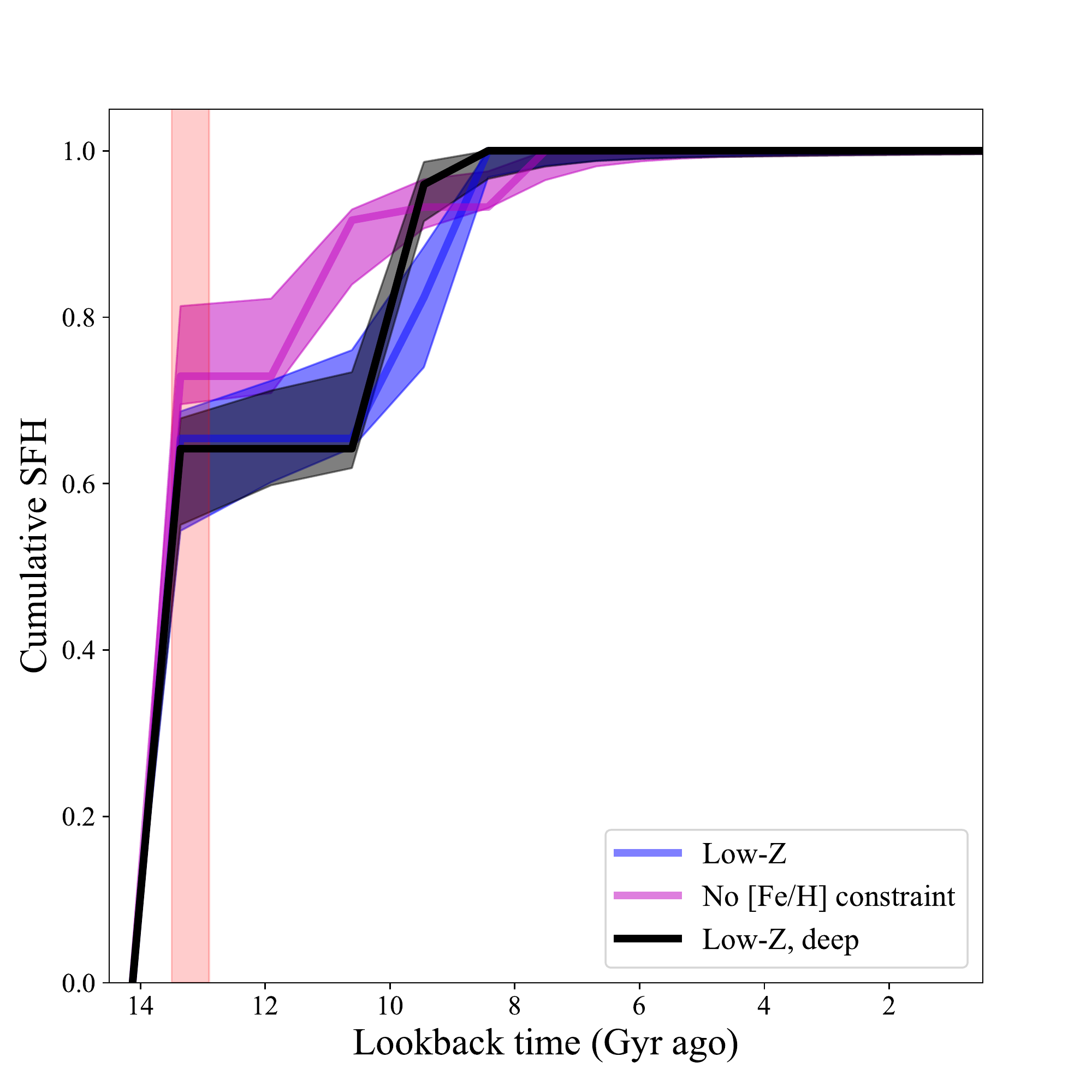}
	\includegraphics[width=\columnwidth]{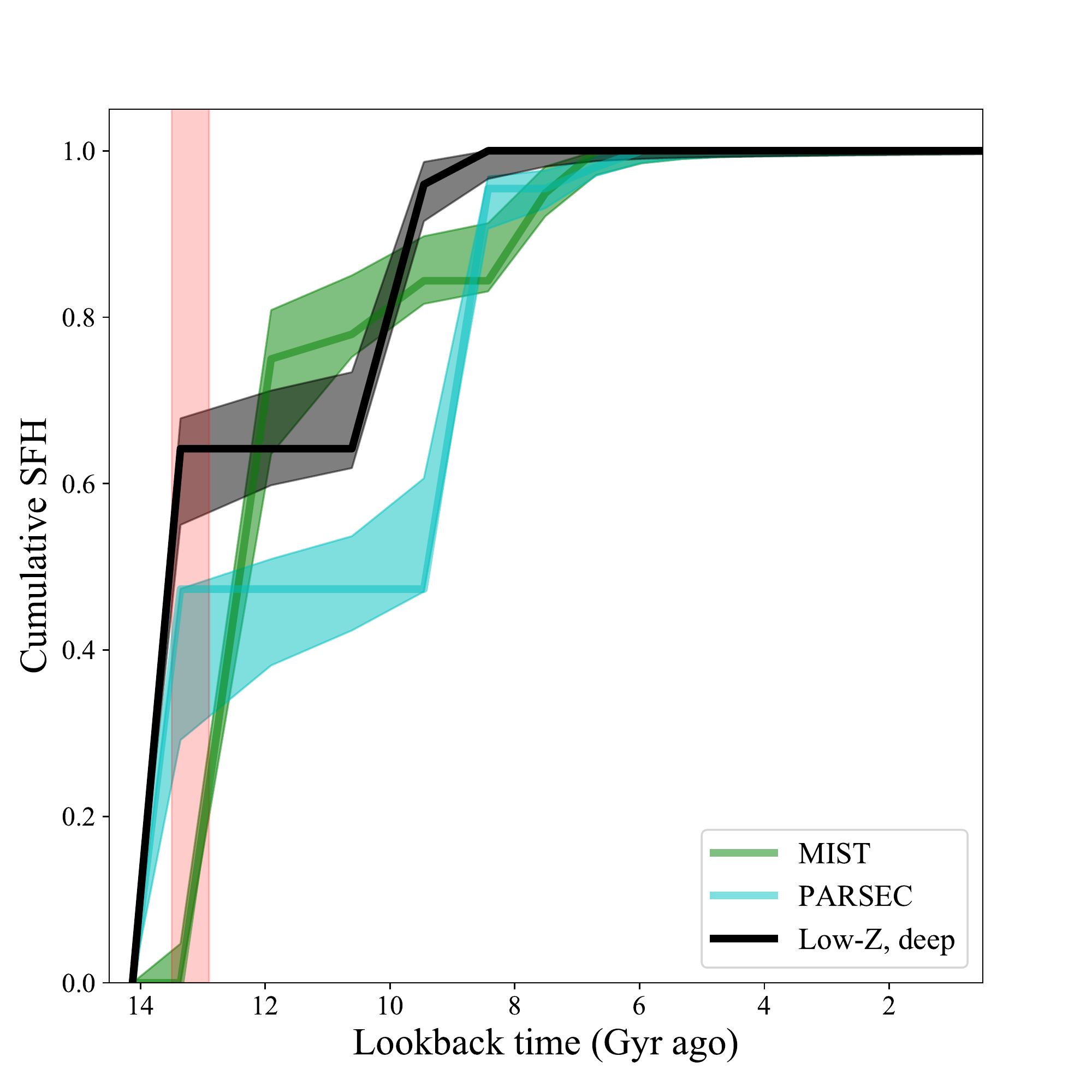}
    \caption{{\bf Left:} The cumulative star formation history of And XIX using MATCH with the Padova stellar evolution models \citep{dolphin02,girardi10}. The solid black line represents our best fit to the data, where we restrict metallicity to [M/H]<-1.7. The grey band shows the 68\% confidence intervals from the random and statistical uncertainties. The magenta line is the same, but without restricting to low metallicities, and the blue curve is obtained by restricting the metallicity and clipping out the faintest 0.3 mag to check we are not biased by noise at the faint-end of the luminosity function. {\bf Right:} The black curve is the same as the left panel. The cyan and green lines show the results when using the PARSEC and MIST stellar libraries with the low metallicity constraint \citep{bressan12,choi16}. In both panels, the red vertical band shows the period of reionisation \citep{fan06}. {\bf In all cases, And~XIX appears to quench between 8-10 Gyr ago, with no evidence of recent star formation.} }
    \label{fig:sfh}
\end{figure*}

To derive the SFH of And~XIX, we use the software package {\sc MATCH} \citep{dolphin02}. This package forward models the CMD of a galaxy to determine its star formation history. A full account of this process can be found in \citet{weisz14a}, and here we provide a summary of the method. Using the Padova \citep{girardi00,marigo08,girardi10}, MIST \citep{choi16}, and PARSEC \citep{bressan12} stellar evolution models, we fit the CMD of And~XIX.  From an inspection of the residuals and overall likelihoods, we find the Padova models provide the best overall fit. The random statistical uncertainties on the SFH are determined using a hybrid Monte Carlo approach \citep{dolphin13}.  We can then estimate systematic uncertainties by fitting multiple independent model sets and comparing the results (which we show in the right hand panel of fig.~\ref{fig:sfh}).  We first run MATCH with no priors on the metallicity of the stellar populations, before restricting it to metallicities of [M/H]$<-1.7$~dex. This cut is based on our previous spectroscopic analysis \citep{collins20}, where we measure And~XIX to be metal poor, with a mean [Fe/H]$=-2.07\pm0.02$. By implementing this prior, we prevent the model from assuming unrealistically metal-rich populations in And XIX. The results of this restriction are broadly consistent with no metallicity restriction, the main difference being that more stars form prior to reionisation in the unrestricted case (making the stellar populations older overall). As such, the metallicity constraint gives a more conservative limit for when And XIX finished forming stars. We show the results from each of these assumption in the left panel of fig.~\ref{fig:sfh}.

As the data are noisy at the deepest limits of our observations, we rerun our low metallicity analysis, clipping out the faintest 0.3 mag above our 90\% completion depths in both bands. This shallower data set still includes And~XIX's oldest MSTO (located at $F814W\sim27.7$), and gives broadly consistent results to our deep analysis, with $>90\%$ of the stellar mass forming prior to 9.5 Gyr ago. 

\section{Results}
\label{sec:results}

In fig.~\ref{fig:sfh}, we show the cumulative SFH of And~XIX for the range of assumptions (left panel, discussed above) and different stellar evolution libraries (right panel). The solid black line in both panels show the results from our deep, low metallicity analysis with the Padova models \citep{girardi10}, with the grey band representing the 68\% confidence intervals from the random plus systematic uncertainties. In the left panel, we show results only from the Padova models, implemented using both the full deep dataset and the low metallicity constraint. The fit with the metallicity prior is the best fit to the data, based on the maximimum likelihood approach. All three SFHs are broadly consistent, showing that $>50\%$ of the stellar mass formed prior to the end of reionisation (red vertical band, \citealt{fan06}), and $>90\%$ forming prior to 9~Gyr ago, with no star formation evident within the past 8~Gyr. This is in sharp contrast to what is observed for similarly luminous dSphs in M31, which we return below.

We chose the Padova models for our SFH, as these produce the highest likelihood values, but there exist a range of stellar evolutionary libraries with which to measure the evolution of And~XIX. The differences between models are typically larger than the statistical uncertainties from using a single library, so it is prudent for us to see how different stellar models modify our findings. In the right panel, we compare the Padova models with the MIST \citep{choi16} and PARSEC \citep{bressan12} libraries. The MIST models produce a similar result to the Padova models, but the break in SF at $t\sim10$~Gyr is less prominent. The PARSEC models push star formation to slightly later times, but And~XIX still ceases forming stars prior to 8~Gyr ago. This trend of PARSEC pushing star formation to slightly later epochs than MIST and Padova is seen in other works, such as \citet{skillman17} in their study of 6 Andromeda dSphs. These differences are likely due to the different treatment of complex stellar physics between models.

Another interesting feature seen across all models is that star formation in And~XIX appears to pause, or slow down, in the 2 Gyr immediately following reionisation (or even later in the MIST models). Could this make And~XIX a candidate for a reionisation fossil? Such fossils are defined as galaxies whose star formation is completely shut down, or ``quenched'' by reionisation \citep{bovill11b}. In the case of And~XIX, the star formation then resumes, requiring a mechanism (such as accretion of new gas) to reinvigorate this process.

To aid our discussion of the nature of And~XIX, we define a quenching time for the galaxy. Following \citet{weisz15} and \citet{skillman17}, we define the epoch of quenching to equal the time at which 90\% of star formation is complete, $\tau_{90}$. Using our fiducial model (solid black lines in fig.~\ref{fig:sfh}), this gives $\tau_{90} = 9.7\pm0.2$~Gyr. We can also measure the time at which 50\% of the stellar mass was in place, $\tau_{50} = 13.5\pm0.1$~Gyr. 

\begin{figure}
	\includegraphics[width=\columnwidth]{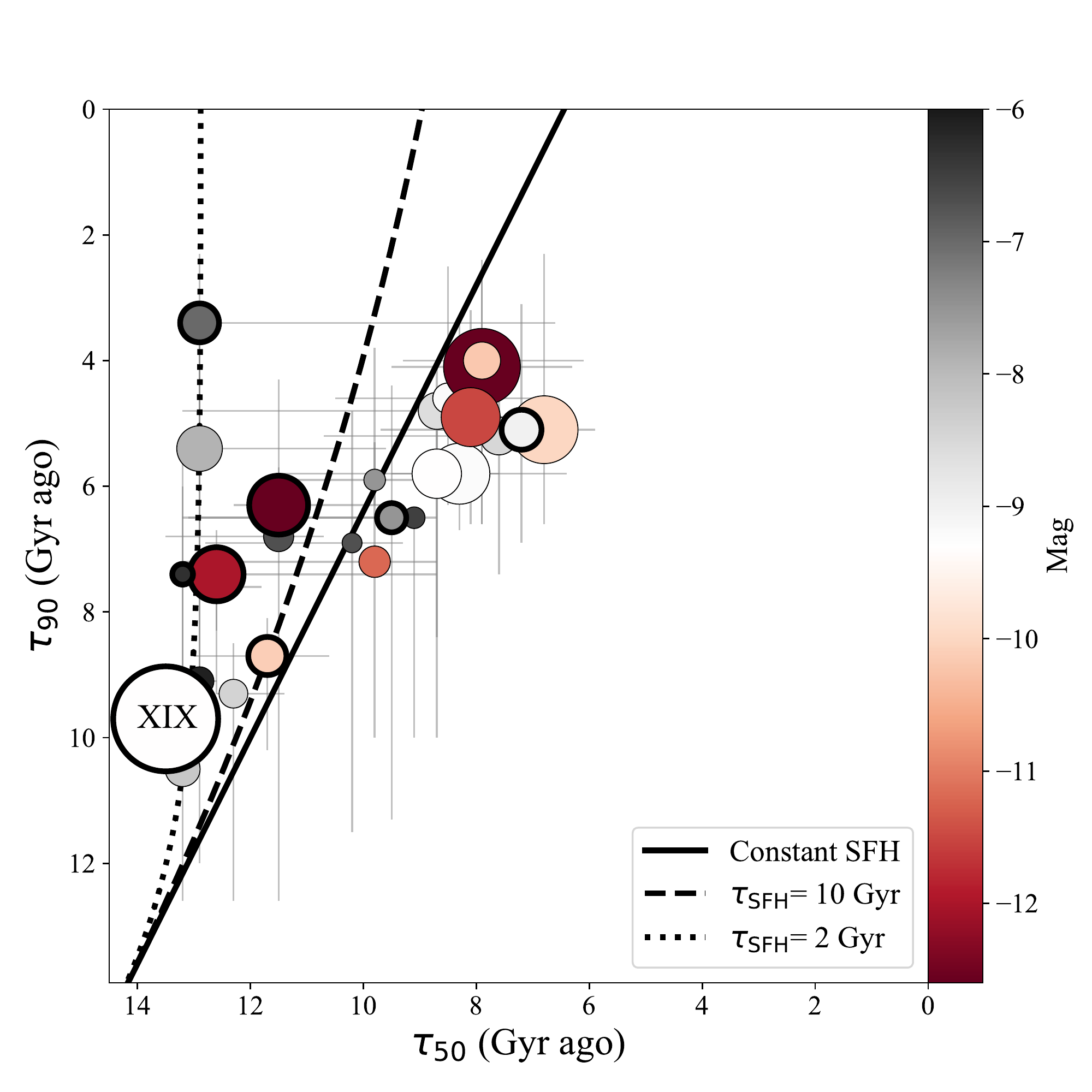}
    \caption{A comparison of the time for 50\% of the stellar mass of M31 dwarf galaxies to form ($\tau_{50}$), vs their quenching time, where 90\% of the stellar mass has formed ($\tau{90}$). Point sizes reflect their half light radii, and the colour shows their luminosity. There are 6 M31 dSphs that have had their SFH measured from similarly deep HST imaging reaching the MSTO (highlighted with black circles, \citealt{skillman17}, and a further 20 for which SFHs have been measured from shallower data reaching just below the horizontal branch \citep{weisz19a}. And~XIX seems to quench much more rapidly than other M31 dSphs of similar luminosity. Could this be due to an early infall, or reionisation?}
    \label{fig:taus}
\end{figure}

To put these results into context, we compare And~XIX with other M31 dSphs for which SFHs have been measured in fig.~\ref{fig:taus}. There are 6 M31 dSphs that have had their SFH measured from similarly deep HST imaging reaching the MSTO (highlighted with black circles, \citealt{skillman17}, and a further 20 for which SFHs have been measured from shallower data reaching just below the horizontal branch \citep{weisz19a}. In fig.~\ref{fig:taus}, we plot $\tau_{50}$ vs. $\tau_{90}$ for the M31 satellite population. The points are colour coded by their luminosity, and the sizes relate to their half light radii. And~XIX is interesting, as it seems to be one of the earliest quenched satellites. It also appears to quench much earlier than similarly bright satellites, most of which continue to form stars for $\sim4$~Gyr longer than And~XIX.

In a wider Local Group context, we can also compare And~XIX to the satellites of the Milky Way. In our own backyard, we tend to see a number of satellites with early quenching times, akin to And~XIX. In particular, Sculptor, Ursa Minor and Draco all quenched $\sim9-10$ Gyr ago, and have comparable stellar mass to And~XIX \citep{weisz15}. These three satellites are thought to have early infall times (comparable to their quenching times). And~XIX may be analogous to these in being among the first to fall into the putative M31 system.

\section{Discussion}
\label{sec:disc}

In this section, we turn our attention to the formation mechanism of And~XIX. Extremely diffuse galaxies are likely formed via a variety of pathways, and a detailed star formation history can permit us to discern between some proposed mechanisms. Ideally, we need a scenario which explains each of the following observations:

\begin{enumerate}
    \item Its large half-light radius of $r_{\rm half}=3357^{+816}_{-465}\,{\rm pc}$ and extremely low surface brightness of $\mu_{V,0}=29.3\pm0.4$ \citep{martin16c}.
    \item Its low density dark matter halo \citep{collins20}.
    \item Its lower than expected metallicity of [Fe/H]$=-2.07\pm0.02$ \citep{collins20}
    \item Its early quenching time, $\tau_{90}=9.7\pm0.2$~Gyr.
\end{enumerate}

\noindent Below, we re-examine several proposed formation mechanisms for And~XIX, and whether they are consistent with these 4 criteria.

\subsection{Bursty star formation as a formation mechanism}

Energetic feedback from star formation has long been discussed as a mechanism for lowering the central mass and density of a galaxy \citep[e.g.][]{navarro96,pontzen12}. During bursty star formation, energy is deposited into both the baryonic and dark matter components of a galaxy. \citet{read19a} have shown how such star formation in dwarf galaxies can `heat' dark matter, gradually transforming a steep central cusp into a lower-density core, and the simulations of \citet[e.g.][]{dicintio17} show that explosive feedback can also turn a typical dwarf galaxy into an ultra diffuse system. Could this explain And~XIX's diffuse nature and low central density?

Given the early quenching time for And~XIX of $\tau_{90}\sim10$ Gyr, it seems unlikely that star formation can explain its unusual properties. The work of \citet{read19a} shows that to truly lower the central density of a dwarf galaxy, prolonged star formation is needed and that galaxies quenched earlier than 6 Gyr ago are unlikely to have been significantly impacted. Similarly, the UDGs formed from star formation feedback in the NIHAO simulations analysed by \citet{dicintio17} all show signs of recent and prolonged star formation. In the case of And~XIX, this can be confidently ruled out. Feedback from bursty star formation can explain neither And XIX's low density dark matter halo, nor its diffuse and extended stellar population.

\subsection{Tidal interaction as a formation mechanism}

Another proposed mechanism for the formation of And~XIX is through tidal processes, specifically impulsive tidal shocking interactions \citep{ogiya18,amorisco19a} and/or tidal stripping \citep{collins13,collins20,jackson21}. In these scenarios, And~XIX would be on a radial orbit, passing close to M31, leading to a significant tidal interaction that causes it to ``puff'' up. Its dark and baryonic matter would be redistributed outwards, causing it to have a lower central density and surface brightness. Such transformations from normal dwarf to UDG are debated. While recent focused $N-$body simulations by \citet{errani21} and \citet{borukhoveskaya21} have shown that tides alone cannot reproduce the extended radii of these galaxies if they are embedded in cusped halos, others find tidal shocking and stripping can explain both their extremely low masses and extended radii \citep[e.g.][]{jiang19,liao19,maccio21,moreno22}. Finally, signs of tidal interactions have been directly observed for several extended UDGs \cite[e.g.][]{jones21}, including NGC 1052-DF4 \citep{montes20}, arguing in favour of the importance of interactions.

Given And~XIX's early quenching $\sim10$~Gyr ago, it is plausible that it had an early infall time onto M31. Any remaining gas would be removed via ram-pressure stripping, consistent with the SFH we measure. This would imply that And~XIX has been exposed to tidal interactions with its host for a long time. However, merely having an early infall time is not enough to form a UDG. In the Milky Way, we see three dSphs with a similar stellar mass and SFH to And~XIX -- Draco, Sculptor and Ursa Minor. They also likely had an early infall onto their host, but with $r_{\rm half}\sim200-250$~pc, none of them are as extended and low surface brightness as And~XIX. 

If And~XIX is on a radial orbit, passing close to its massive host, this could explain its low surface brightness, extended half-light radius, the hints of tidal stripping seen in its outskirts in PAndAS imaging \citep{mcconnachie08, martin16c} and its dynamics \citep{collins20}. \citet{watkins13} perform a statistical analysis on the orbital properties of Andromeda satellites, using the timing argument and phase-space distribution functions. They find it most likely that And~XIX is not on a highly radial orbit, and that it has an orbital pericentre of $r_p=112\pm12$~kpc, i.e. only slightly smaller than its current distance of $D_{M31}=115^{+96}_{-9}$~kpc. Taken at face value, it would be unlikely that such a large pericentre could result in a strong enough tidal interaction to puff up And~XIX.  It could be that And~XIX was tidally affected by an interaction with another system prior to infall, as recent work by \citet{genina20} showed that a significant fraction of tidally stripped Fornax-like dwarf galaxies in the APOSTLES simulations begin to lose mass prior to infall as a result of interactions with other systems. 

Without proper motions, we cannot determine the true orbit of And~XIX. However, we can use some of its other properties to constrain whether tidal stripping may have occurred. If And~XIX has experienced significant stellar mass loss, we may expect it to have a higher stellar metallicity than other dwarf galaxies of similar brightness. But we in fact see the opposite. In fig.~\ref{fig:lfeh}, we show the luminosity-metallicity relation for dSphs galaxies \citep{kirby13b}, and And~XIX is highlighted as a magenta star, sitting below this relation with far smaller uncertainties than other outliers in the same region of parameter space. If tidal stripping (as opposed to tidal shocking) has occurred, it likely has not resulted in much stellar mass loss. As dwarf galaxies lose $\sim90\%$ of their halo mass before significant stripping of the stars even begins (e.g. \citealt{penarrubia08b}), And~XIX could have undergone significant stripping without moving in the stellar mass-metallicity plane. Finally, tidal shocking or stirring could still have affected the dynamics and structural properties of And~XIX without any significant stellar mass loss. Future proper motion measurements and dedicated orbital modelling of And~XIX would allow us to constrain its orbital history, and better comment on the tidal stripping, stirring and shocking scenarios.

\subsection{Mergers as a formation mechanism}

\begin{figure}
     \includegraphics[angle=0,width=\columnwidth]{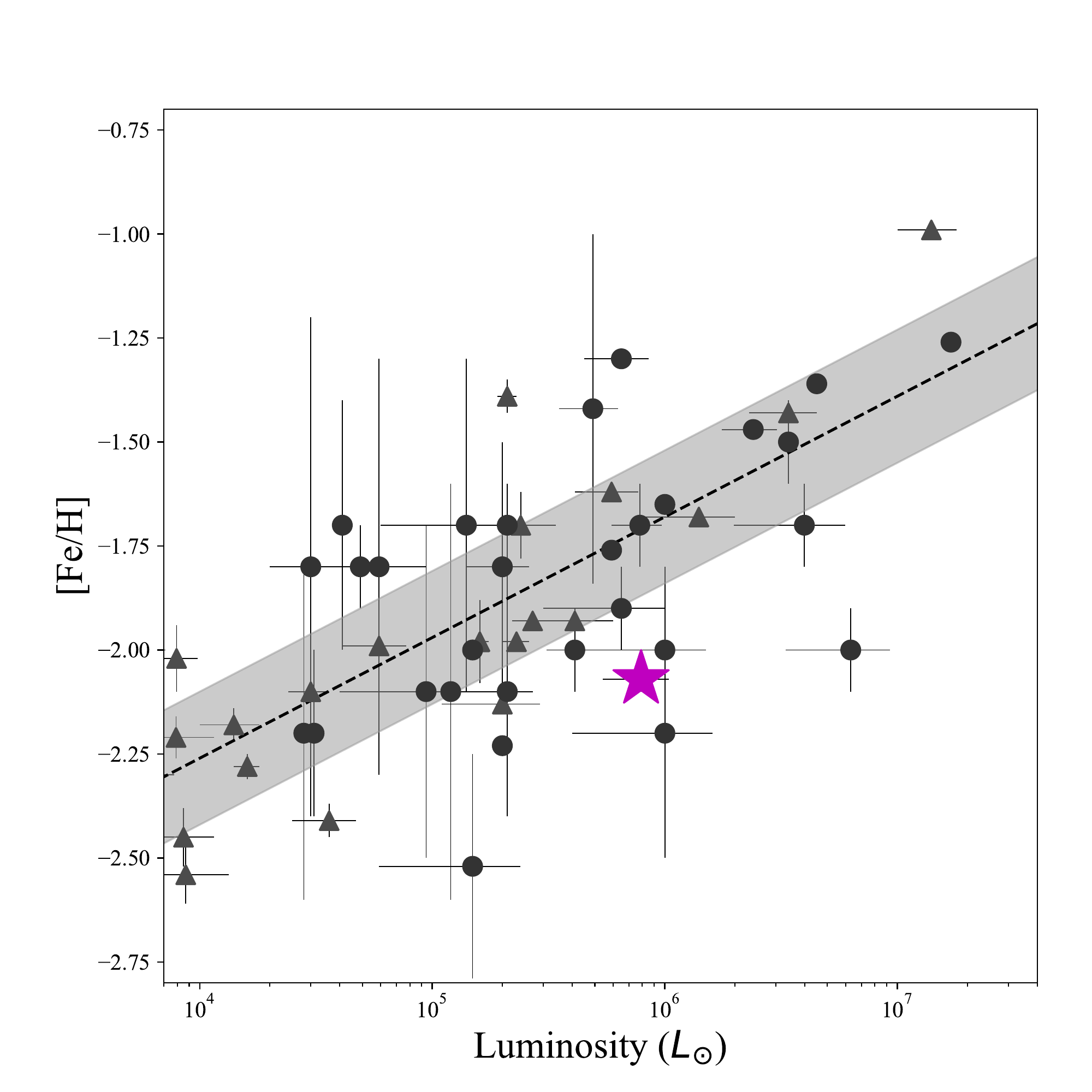}
    \caption{Here we show the metallicity for And~XIX (magenta star \citealt{collins20}) in comparison with other Local Group dSphs. The luminosity vs. [Fe/H] relation for MW dSphs is shown with the dashed line, and the grey shaded region shows the $1\sigma$ scatter. The triangles show the Milky Way data, while M31 dSphs are shown with circles \citep{collins13,martin14b, ho15,wojno20a}. And~XIX has a lower metallicity than would be expected based on the luminosity relation, consistent with a merger origin. }
    \label{fig:lfeh}
\end{figure}

Another recent idea for the formation of UDGs is through mergers. In \citet{silk19}, they propose that early (more than 10 Gyr ago) high-velocity collisions between gas rich dwarf galaxies may produce UDGs with little to no dark matter and a healthy population of globular clusters (GCs) like NGC1052-DF2 \citep{vandokkum18}. In this scenario, one would expect a burst of star formation immediately after the collision at a look-back time of $\sim10$~Gyr, which is consistent with our measured SFH. However, it is also expected that such a collision would form a number of bright GCs, similar to $\omega$-Centauri. It is unclear how many GCs would be expected in a galaxy with the same stellar mass as And~XIX, but at present none have been detected. Future work would be needed to calculate the expected number of GCs to form for And~XIX.

Similarly, a late merger scenario has been reported as a channel for forming UDGs in the EDGE hydrodynamical simulations \citep{rey19}. In their work, they perform zoom-in simulations of a single isolated dwarf galaxy selected from their larger cosmological volume with slightly differing initial conditions. They re-scale the initial density of the dwarf galaxy, resulting in a range of assembly times. Those with a higher density assemble their stellar mass earlier, while those with lower density build theirs later at lower redshift. The lowest density (and hence, latest assembly time) dwarf is quenched rapidly by reionisation, and then late mergers build the the bulk of its stellar mass (approximately 94\%) at $z<6$ (look-back time of less than 12.8 Gyr). These stars are preferentially deposited at large radii leading to an extended ($r_{\rm half}=820$~pc), low surface brightness dwarf. This specific merger history, where the stellar mass builds up later through mergers, lies within the 68\% contour of assembly histories for the parent cosmological EDGE simulation of 1500 systems, meaning that late dry mergers like this are not uncommon. And while their final UDG is both fainter and smaller than And~XIX, it does show a new path for UDG formation. It is unclear from this single example if any of these mergers could fuel a burst of star formation in the primary as we see for And XIX, but if any brought in gas they could reignite star formation.

Additionally, \citet{rey19} show that such UDGs should be more metal-poor on average than similarly luminous galaxies. This is precisely what we see for And~XIX, as shown in fig.~\ref{fig:lfeh}. This low metallicity is quite intriguing as while a higher metallicity may be expected for a system undergoing significant stellar mass loss, it’s not clear what causes a lower metallicity other than very strong, bursty feedback \cite[e.g.][]{prgomet22}, which is inconsistent with our measured SFH. In this scenario, the lower metallicity would result from And~XIX forming as a merger product of multiple ``fossil'' galaxies, whose star formation were largely quenched as a result of reionisation. In the FIRE simulations, galaxies with stellar mass below $M_*\sim10^5\,{\rm M}_\odot$ are always quenched by reionisation \citep{wheeler19}. Given And XIX's stellar mass of $M_*=7.9\times10^5\,{\rm M}_\odot$, it would require a number of late mergers to assemble solely from reionisation fossils depositing their mass at large radii, just as the ultra-faint dwarf in \citet{rey19}, but at a higher stellar mass.

The merger pathway is an intriguing formation scenario, as it would explain And~XIX's size, surface brightness, early quenching and low metallicity. Given that all the merged systems would form the bulk of their stars prior to reionisation, it may be very challenging to disentangle the progenitors with current data and either rule out or confirm this hypothesis. Detailed, precision abundances for its constituent stars may help, but such measurements are currently extremely challenging to make. However, additional simulation and modelling aimed at objects like And XIX might make it clearer what 'smoking guns' to search for in this intriguing system.

\section{Conclusions}
\label{sec:conclusion}

In this work, we present a quantitative star formation history for And~XIX, derived from deep HST imaging which resolves the oMSTO of the dwarf galaxy. We report the following findings:

\begin{itemize}
    %\item We update the distance to And~XIX using the magnitude of its HB population, finding $D=746\pm22$~kpc, consistent with previous work. Using this precise distance, we update the physical size of And~XIX to $r_{\rm half}=3081^{+743}_{-422}\,{\rm pc}$ and its absolute magnitude to $M_V=-9.9\pm0.4$.
    
    \item Using MATCH and the Padova stellar evolution libraries, we measure a detailed SFH for And~XIX, finding that it formed the bulk of its stellar mass early. We find that there is no evidence of any star formation in the past 8 Gyr (a result that is independent of stellar evolution model used).
    
    \item We measure a quenching time for And~XIX -- defined as the point where 90\% of star formation is complete -- finding $\tau_{90}=9.7\pm0.2$~Gyr. This is earlier than similarly bright M31 dSphs for which detailed SFHs have been measured, however it is consistent with the comparably luminous MW satellites, Draco, Sculptor and Ursa Minor. These are all candidates for early infall onto the Galaxy, implying And~XIX may have been accreted early to M31.
    
    \item With the addition of a detailed SFH, we re-examine several mechanisms proposed for the formation of And~XIX. We can rule out that its low density halo and diffuse, extended stellar populations can be explained with feedback from bursty star formation given the early quenching time. Tidal stripping could explain its structure and density, but not its low metallicity. Tidal stirring and/or shocking may still have played a role in altering its stellar populations without affecting its stellar mass. A late dry merger scenario -- similar to that presented in \citet{rey19} -- could explain all of its present day properties, including its metallicity. A further epoch of HST imaging would allow us to constrain its proper motion and test the tidal scenario. Detailed abundances and follow-up modelling could allow us to determine whether mergers are important in the system.
    
\end{itemize}

And~XIX remains a fascinating diffuse galaxy, and presents us with an opportunity to better understand how the lowest surface brightness galaxies form. With near-future observations, we can soon confidently understand whether this system formed through tidal processes or as the result of mergers.

\section*{Acknowledgements}

We thank the anonymous reviewer for their helpful and insihtful comments. 

EB acknowledges support from a Vici grant from the Netherlands Organisation for
Scientific Research (NWO).

This research is based on observations made with the NASA/ESA Hubble Space Telescope obtained from the Space Telescope Science Institute, which is operated by the Association of Universities for Research in Astronomy, Inc., under NASA contract NAS 5–26555. These observations are associated with program ID 15302; support for this work (BFW, KMG, and AD) was provided by NASA through grant HST-GO-15302.

\section*{Data Availability}

The HST data used in this work are publicly available on the Barbara A. Mikulski Archive for Space Telescopes (MAST \href{https://mast.stsci.edu/portal/Mashup/Clients/Mast/Portal.html}{https://mast.stsci.edu/portal/Mashup/Clients/Mast/Portal.html}). The source catalogues of resolved photometry used in this work will be made available upon reasonable request to the lead author.

%%%%%%%%%%%%%%%%%%%%%%%%%%%%%%%%%%%%%%%%%%%%%%%%%%



\bibliographystyle{mnras}
\bibliography{michelle} % if your bibtex file is called example.bib

%\begin{thebibliography}{99}
%\end{thebibliography}

%%%%%%%%%%%%%%%%%%%%%%%%%%%%%%%%%%%%%%%%%%%%%%%%%%

%%%%%%%%%%%%%%%%% APPENDICES %%%%%%%%%%%%%%%%%%%%%

\appendix

\section{Tabulated star formation histories}

Here we present the full star formation history results from all models plotted in fig.~\ref{fig:sfh}. The first three columns contain the results from the Padova models with now constraints, the metallicity prior with a magnitude limit of 0.3 magnitudes above the 90\% completeness limit (Low-Z), and the metallicity prior without the magnitude limit (Low-Z deep). The MIST and PARSEC results include the metallicity prior, and no magnitude limit (the same as Low-z deep). 

\begin{table*}
\caption{Cumulative SFH as a function of lookback time measured using different models and priors.}
\begin{tabular}{cccccc}
\hline
Cumulative & Padova & Low-Z & Low-Z-deep & MIST & PARSEC  \\
SFH (\%) & (Gyr) & (Gyr) & (Gyr) & (Gyr) & (Gyr)  \\
\hline
\hline
10 & 14.13$^{+0.00}_{-0.00}$ & 14.13$^{+0.00}_{-0.00}$ & 14.13$^{+0.00}_{-0.00}$ & 13.36$^{+0.77}_{-0.00}$ & 14.13$^{+0.00}_{-0.00}$ \\
20 & 14.02$^{+0.01}_{-0.01}$ & 14.01$^{+0.01}_{-0.02}$ & 14.01$^{+0.01}_{-0.02}$ & 13.16$^{+0.09}_{-0.03}$ & 13.96$^{+0.00}_{-0.10}$ \\
30 & 13.91$^{+0.02}_{-0.01}$ & 13.89$^{+0.01}_{-0.05 }$& 13.89$^{+0.01}_{-0.04}$ & 12.97$^{+0.10}_{-0.07 }$& 13.80$^{+0.00}_{-0.20}$ \\
40 & 13.81$^{+0.03}_{-0.02}$ & 13.77$^{+0.02}_{-0.07}$ & 13.77$^{+0.02}_{-0.06}$ & 12.78$^{+0.10}_{-0.10}$ & 13.64$^{+0.00}_{-0.41}$ \\
50 & 13.70$^{+0.04}_{-0.02}$ & 13.66$^{+0.02}_{-0.10}$ & 13.65$^{+0.03}_{-0.08}$ & 12.58$^{+0.10}_{-0.14}$ & 13.48$^{+0.00}_{-2.14}$ \\
60 & 13.60$^{+0.05}_{-0.03}$ & 13.54$^{+0.03}_{-0.12}$ & 13.53$^{+0.03}_{-0.10}$ & 12.39$^{+0.10}_{-0.17 }$& 9.40$^{+2.88}_{-0.01}$ \\
70 & 13.49$^{+0.07}_{-0.03}$ & 13.42$^{+0.03}_{-1.47}$ & 13.41$^{+0.04}_{-1.63}$ & 12.19$^{+0.11}_{-0.21}$ & 9.19$^{+0.38}_{-0.04}$ \\
80 & 13.39$^{+0.08}_{-0.54}$ & 10.30$^{+2.54}_{-0.37}$& 10.40$^{+2.02 }_{-0.11}$ & 12.00$^{+0.11}_{-0.81}$ & 8.97$^{+0.22}_{-0.06}$ \\
90 & 11.42$^{+1.95}_{-0.42}$ & 9.62$^{+0.62}_{-0.43}$ & 10.04$^{+0.27}_{-0.13}$ & 10.24$^{+1.68}_{-0.50}$ & 8.76$^{+0.15}_{-0.08}$ \\
100 & 10.72$^{+0.24}_{-1.15}$ & 9.46$^{+0.00}_{-0.72}$ & 9.67$^{+0.18}_{-0.16}$ & 7.94$^{+1.34}_{-0.21}$ & 8.54$^{+0.08}_{-0.10}$ \\
\hline
\end{tabular}
\end{table*}


%%%%%%%%%%%%%%%%%%%%%%%%%%%%%%%%%%%%%%%%%%%%%%%%%%

% Don't change these lines
\bsp	% typesetting comment
\label{lastpage}
\end{document}